\documentclass[pre,twocolumn,superscriptaddress,showpacs]{revtex4-1}
\newcommand{\revision}[1]{{{#1}}}

\usepackage[dvipdfmx]{graphicx}
\usepackage{bmpsize}
\usepackage{dcolumn}% Align table columns on decimal point
\usepackage{bm}% bold math
\usepackage{amsmath, amssymb}
\usepackage{feynmp}
\usepackage{url}
\usepackage{setspace}
\usepackage{color}
\usepackage{hyperref}
\usepackage{ulem}

\newcommand{\beq}{\begin{equation}}
\newcommand{\eeq}{\end{equation}}
\newcommand{\beqa}{\begin{eqnarray}}
\newcommand{\eeqa}{\end{eqnarray}}
\newcommand{\Tr}{\text{Tr}}

\newcommand{\calQ}{\mathcal{Q}}
\newcommand{\calL}{\mathcal{L}}

\newcommand{\gamu}{\Gamma^{+}}
\newcommand{\gamd}{\Gamma^{-}}

\begin{document}

\title{Dynamics of a quantum system interacting with \revision{white} non-Gaussian baths: Poisson noise master equation}

\author{Ken Funo}
\affiliation{Institute for Molecular Science, National Institutes of Natural Sciences, Okazaki 444-8585, Japan}
\email{kenfuno@ims.ac.jp}

\author{Akihito Ishizaki}
\affiliation{Institute for Molecular Science, National Institutes of Natural Sciences, Okazaki 444-8585, Japan}
\affiliation{Graduate Institute for Advanced Studies, SOKENDAI, Okazaki 444-8585, Japan}
\email{ishizaki@ims.ac.jp}

\date{\today}

\begin{abstract}
Quantum systems are unavoidably open to their surrounding degrees of freedom. 
The theory of open quantum systems is thus crucial to understanding the fluctuations, dissipation, and decoherence of a quantum system of interest. 
Typically, the bath is modeled as an ensemble of harmonic oscillators, which yields Gaussian statistics of the bath influence on the quantum systems. 
However, there are also phenomena in which the bath consists of two-state systems, spins, or anharmonic oscillators; therefore, the non-Gaussian properties of the bath become important. 
Nevertheless, a theoretical framework to describe quantum systems under the influence of such non-Gaussian baths is not well established. 
Here, we develop a theory to describe quantum dissipative systems affected by Poisson noise properties of the bath, because the L\'evi-It\^o decomposition theorem asserts that Poisson noise is fundamental in describing arbitrary white noise beyond Gaussian properties.  
We introduce a quantum bath model that allows for the consistent description of dissipative quantum systems. The obtained master equation reveals non-Gaussian bath effects in the white noise regime, and provides an essential step toward describing open quantum dynamics under the influence of generic baths. 
\end{abstract}

\maketitle

{\it Introduction.---} 
In principle, quantum systems cannot be regarded as isolated systems because they are unavoidably in contact with their outside world; hence, their quantum natures are sometimes sustained and sometimes destroyed. Especially in condensed matter systems such as solids and complex molecules, quantum systems are affected by numerous dynamic degrees of freedom. The balance between robustness and fragility of the quantum natures may dramatically alter the properties and dynamical behaviors of these systems. Therefore, it is crucial to investigate the impact of ambient degrees of freedom on quantum systems in various research fields, ranging from condensed matter physics~\cite{Weiss, nonHermitian} to quantum biophysics~\cite{renger01, ishizaki10, quantum_effects_biology, Chin12}. Quantum information science, which has been making remarkable progress in recent years, is no exception~\cite{decoherence_free1, decoherence_free2, noiseless, Cirac09, Gardiner, nonMarkov, Deffner}. In quantum information science, it is crucial to elucidate and eradicate noise sources to achieve long quantum coherence times, which is necessary for practical applications in quantum computing, communications, and sensing.

In the literature of such investigations, the surrounding degrees of freedom affecting a quantum system under study are referred to as {\it bath} or {\it environment}. Typically, the bath has been modeled as an ensemble of harmonic oscillators, namely a boson bath~\cite{Caldeira83}, which introduces Gaussian fluctuations and dissipation into the quantum system. However, these Gaussian baths cannot characterize noise sources, such as the shot noise in mesoscopic conductors \cite{Shotnoise1, Shotnoise2} and radiation pressure inside optomechanical systems \cite{Shotnoise3}. 
It is also believed that one of the primary sources of decoherence in solid-state quantum devices could be two-state fluctuators, which represent the ambient microscopic degrees of freedom~\cite{1fnoise, twolevelbath1, twolevelbath2, twolevelbath3}. 
\revision{There is significant interest in utilizing dissipative qubits to model and implement tunable baths~\cite{tunable0, tunable1, tunable2, tunable3, tunable4, tunable5}, with possible applications to dissipative quantum state engineering~\cite{Cirac09, tunable4}. }
Moreover, anharmonicity in lattice vibrations has been argued to increase charge carrier lifetimes in the lead halide perovskite materials~\cite{perovskite}. However, a theoretical framework describing the dynamics of quantum systems under the influence of such non-Gaussian baths is not well established~\cite{Pekola23, Hanggi, spinbath, Strunz17, Strunz23, Cao1, Cao2}. 

To develop a theory to describe the impact of a non-Gaussian bath on a quantum system, we address the L\'evi-It\^o decomposition theorem of stochastic processes~\cite{Applebaum, Gardiner09, Kanazawa}. This theorem states that any white noise, in which the noise correlation time is negligibly short, decomposes into Gaussian white noise and a sum of Poisson white noises. Therefore, describing quantum dissipative systems under the influence of Poisson noise facilitates the investigation of the effect of generic baths with arbitrary white noise properties.

In this work, we develop a theory to describe quantum dissipative systems affected by Poisson noise and discuss the physical properties of the resulting Poisson bath master equation. 
\revision{To explore the noise properties of the bath, past studies have primarily focused on cases where the bath is modeled as a classical stochastic noise source~\cite{PoissonME1, PoissonME2, PoissonME3, Mukamel78, Kosloff, classicalnoise2, Kubo, classicalnoise1}. On the other hand, we start with a quantum mechanical modeling of the bath using dissipative two-level systems. It is worth pointing out that our setup is relevant to physical situations of two-state fluctuators~\cite{1fnoise, twolevelbath1, twolevelbath2, twolevelbath3} and engineered bath~\cite{tunable0, tunable1, tunable2, tunable3, tunable4, tunable5}.} 
We demonstrate that the property of the constructed bath model is consistent with the Poisson noise statistics when the bath correlation time is short and the bath interacts with the quantum system strongly but discretely. We then derive an equation of motion to describe the time evolution of a quantum system coupled to the constructed Poisson bath. 
The statistical differences between the Poisson and Gaussian baths are captured in the emission and absorption rates. The rates of the obtained Poisson noise master equation saturate as the system-bath coupling increases, whereas those of the Gaussian noise master equation depend quadratically on the coupling. These differences are illustrated by considering a Dicke super-radiance model, which describes a collective coupling between identical two-level systems and a bath~\cite{Dicke}.

%%%%%%%%%%%%%%%%%%%%%%%%%%%%%%%%%%%%%%%%%%%%%%%%%%%%%%%%%%%%%%%%%%%%%%%%%%%%%%%%%%%
{\it System-bath model.---} 
We consider a situation where a general system is coupled to a bath whose inverse temperature is $\beta$. 
The time-evolution of the entire system is described by the Liouville equation,
\beq
\partial_{t}\rho_{\rm SB}(t) = \mathcal{L}\rho_{\rm SB}(t), 
\label{Eq1}
\eeq
where $\mathcal{L}=\calL_{\rm S}+\calL_{\rm B}+\calL_{\rm int}$ is the Liouville superoperator defined by $\mathcal{L}_{\rm S}\bullet = -i[H_{\rm S},\bullet]$ and $\mathcal{L}_{\rm int}\bullet=-i[H_{\rm int},\bullet]$ describes the effect of unitary time-evolution according to the system Hamiltonian $H_{\rm S}$ and interaction Hamiltonian $H_{\rm int}$, respectively (we set $\hbar=1$). 
In the standard approach~\cite{Freed75}, $\calL_{\rm B}$ is defined as $-i[H_{\rm B},\bullet]$, where $H_{\rm B}$ is the bath Hamiltonian. We could instead consider a non-unitary time-evolution modeled by the Gorini-Kossakowski-Sudarshan-Lindblad (GKSL) master equation $\mathcal{L}_{\rm B}\bullet=-i[H_{\rm B},\bullet]+\mathcal{D}[\bullet]$~\cite{Lindblad, GKS, Breuer}. Here, $\mathcal{D}$ is the dissipator that causes $B$ to approach the Gibbs state $\rho_{\rm B}^{\rm eq}=\exp(-\beta H_{\rm B})/\Tr[\exp(-\beta H_{\rm B})]$ in the long-time limit, i.e., $\mathcal{L}_{\rm B}\rho_{\rm B}^{\rm eq}=0$. For example, See Eq.~\eqref{diss}. 
In the first approach, the size of the bath is typically assumed to be sufficiently large to avoid recurrence and describe dissipation of the system. In the second approach, the system interacts with a few bath modes denoted by $B$, and the rest of the bath modes are modeled explicitly to induce dissipation via the dissipator. 

{\it Nakajima-Zwanzig projection operator method.---} 
In what follows, we obtain the equation of motion for the reduced density operator of the system, $\rho_{\rm S}(t)=\Tr_{\rm B}[\rho_{\rm SB}(t)]$. To proceed, we introduce the projection operator to the bath Gibbs state, $\mathcal{P}=\rho_{\rm B}^{\rm eq}\Tr_{\rm B}$~\cite{Nakajima, Zwanzig}.  
We also introduce the orthogonal complement, $\mathcal{Q}=1-\mathcal{P}$. The conventional projection operator method considers the case of $\mathcal{D}=0$; however, it can be directly generalized to a case of $\mathcal{D}\neq 0$ as follows. 

We assume that the initial state is given by a product state and the bath is initially in the Gibbs state, $\rho_{\rm SB}(0)=\rho_{\rm S}(0)\otimes \rho_{\rm B}^{\rm eq}$. Additionally, we assume that  
$\Tr_{\rm B}[H_{\rm int}\rho_{\rm B}^{\rm eq}]=0$, which can be always satisfied by modifying the Hamiltonian appropriately. The time-evolution equation for the reduced density operator $\rho_{\rm S}(t)$ is obtained as follows:
\beq
\partial_{t}\rho_{\rm S}(t) 
= 
\calL_{\rm S}\rho_{\rm S}(t) 
+ 
\int^{t}_{0}du \, 
\Tr_{\rm B}
\Bigl[ 
    \calL_{\rm int}
    e^{\calQ\calL\calQ u}
    \calQ
    \calL_{\rm int}
    \rho_{\rm B}^{\rm eq} 
\Bigr] 
\rho_{\rm S}(t-u). 
\label{rhoS}
\eeq
The interaction Hamiltonian can be written as $H_{\rm int}=\sum_{k}S^{k}\otimes B^{k}$, where $S^{k}$ and $B^{k}$ are the system and bath operators, respectively. Additionally, we use the left-right superoperator notation $A_{+}\rho:=A\rho$ and $A_{-}\rho:=\rho A$, and write the Liouville superoperator as $\calL_{\rm int}= \sum_{l=\pm,k}(-i)^{l} S^{k}_{l}B^{k}_{l}$. We then move to the interaction picture defined by $\rho_{\rm S}^{\rm I}(t)=e^{-\calL_{\rm S}t}\rho_{\rm S}(t)$ and $S(t)=e^{-\calL_{\rm S}t}Se^{\calL_{\rm S}t}$ for system operators, and expand Eq.~(\ref{rhoS}) to infinite orders of coupling~\cite{Freed75} to obtain 
\begin{widetext}
\beqa
\dot{\rho}_{\rm S}^{\rm I}(t) 
&=& 
\sum_{n=1}^{\infty} 
\int^{t}_{0}du_{1} \cdots \int^{u_{n-1}}_{0} du_{n} \sum_{\vec{l},\vec{k}} \prod_{j=1}^{n+1}(-i)^{l_{j}} \chi_{n+1}^{\vec{l},\vec{k}}(u_{1},u_{2},\cdots,u_{n},0) 
\nonumber \\
& &\times 
S^{k_{1}}_{l_{1}}(t)
S^{k_{2}}_{l_{2}}(t-u_{1}+u_{2})
%S^{-k_{3}}_{l_{3}}(t-u_{1}+u_{3}) 
\cdots 
S^{k_{n}}_{l_{n}}(t-u_{1}+u_{n})
S^{k_{n+1}}_{l_{n+1}}(t-u_{1})
\rho_{\rm S}^{\rm I}(t-u_{1}),  
\label{Seq}
\eeqa
%where $l =  \sum_{i}^{n+1} l_{i}$ 
where $\chi^{ \vec{l},\vec{k} }_{n}(t_{1},\cdots,t_{n})$ defines the $n$-point bath correlation function with $t_{1}\geq \cdots \geq t_{n}$ as
\beq
\chi^{ \vec{l},\vec{k} }_{n}(t_{1},\cdots,t_{n})
= 
\Tr_{\rm B}
\left[ 
B^{k_{1}}_{l_{1}}
e^{\mathcal{L}_{\rm B}(t_{1}-t_{2})}
\calQ 
B^{k_{2}}_{l_{2}} 
e^{\mathcal{L}_{\rm B}(t_{2}-t_{3})}
\calQ 
B^{k_{3}}_{l_{3}} 
\cdots 
e^{\mathcal{L}_{\rm B}(t_{n-1}-t_{n}) } 
\calQ 
B^{k_{n}}_{l_{n}} 
\rho_{\rm B}^{\rm eq}
\right].
\label{bathC}
\eeq 
\end{widetext}
The mathematical structures of Eqs.~\eqref{Seq} and~\eqref{bathC} serve as a starting point to analyze and classify the influence of non-Gaussian baths on the system. 
What should be noted here is that the effect of general baths on the reduced dynamics of the system is fully captured by the $n$-point bath correlation functions, as shown in Eq.~(\ref{Seq}). Consequently, the systems of two different system-bath models with the same $n$-point bath correlation functions exhibit equivalent reduced dynamics, although their bath properties $\calL_{\rm B}$ and $B^{k}$ may differ.

{\it Random telegraph noise bath model.---} 
We construct an explicit non-Gaussian bath model. In doing so, we view the bath as a quantum noise source and use the properties of random telegraph noise and Poisson white noise, which are renowned examples of non-Gaussian noise. Random telegraph noise takes two values that randomly switch from one to another based on the jump rates~\cite{DMN}. As discussed in Ref.~\cite{VanDenBroeck}, the random telegraph noise converges to Poisson white noise defined by $\xi(t)=\sum_{i}^{n(t)}b_{i}\delta(t-t_{i})$ in the appropriate limit. The probability of having $n$ random time points $t_{i}$ is given by the Poisson distribution $(\gamma t)^{n}e^{-\gamma t}/n!$, the noise strength $b_{i}$ may take random values at each time point following the distribution $p(b)$, and $\gamma$ is the average noise rate. 
%The cumulants are given by $\gamma b^{n}\delta(t_{1}-t_{2})\cdots \delta(t_{n-1}-t_{n})$. 
Poisson white noise converges to Gaussian white noise in the limit where the noise intensity is weak $b\rightarrow 0$ and noise occurs continuously $\gamma\rightarrow \infty$ at constant $\gamma b^{2}$. 
%characterized by its second cumulant $\gamma b^{2}\delta(t_{1}-t_{2})$. 

\revision{Modeling the bath as a classical noise source that fluctuates the system Hamiltonian parameters (e.g., the system frequency) has been investigated~\cite{Kubo, PoissonME1, PoissonME2, PoissonME3, Mukamel78, Kosloff, classicalnoise1, classicalnoise2}. On the other hand, we describe dissipation associated with the random telegraph noise by assuming that $B$ is modeled by a dissipative two-qubit system}
\beq
\calL_{\rm B}\bullet =-i[H_{\rm B},\bullet]+\sum_{i=1,2}\Bigr( \Gamma^{+}_{i}\mathcal{D}_{\sigma^{+}_{i}}[\bullet]+\gamd_{i}\mathcal{D}_{\sigma^{-}_{i}}[\bullet] \Bigr), \label{LE}
\eeq
where $H_{\rm B}=\sum_{i=1,2}(\omega_{i}/2)\sigma^{z}_{i}$ is the bath Hamiltonian, $\Gamma_{i}^{\pm}$ are the jump rates, and $\mathcal{D}_{A}[\bullet]$ is the dissipator written as \cite{Breuer}
\beq
\mathcal{D}_{A}[\bullet] = A \bullet A^{\dagger} -\frac{1}{2}\{ A^{\dagger}A,\bullet \}.  \label{diss}
\eeq 
We assume the detailed balance condition $\gamd_{i}/\gamu_{i}=e^{\beta\omega_{i}}$, which is a sufficient condition for $\calL_{\rm B}\rho_{\rm B}^{\rm eq}=0$ to hold. 
We choose $S^{-}=L, S^{+}=L^{\dagger}$, and $B^{\pm}=\lambda\sigma_{1}^{\mp}\sigma_{2}^{\pm}$, where $\lambda$ is the coupling strength and $L$ is a general (non-Hermitian) system operator. Therefore, the interaction Hamiltonian reads $H_{\rm int} =\sum_{k=\pm}S^{k}B^{k}= \lambda ( L \sigma^{+}_{1}\sigma^{-}_{2} + L^{\dagger}\sigma^{-}_{1}\sigma^{+}_{2})$.

{\it Multiple jump effects.---} 
We first give intuitions on how the bath affects the system in the case of $\gamd_{1}=\gamd_{2}\gg \gamu_{i}$ because this parameter regime is relevant in the Poisson white noise limit.  
Because $\gamd_{1}$ is large, $B$ mostly remains in the ground state $|g_{1},g_{2}\rangle$, and the first spin gets excited based on the rate $\gamu_{1}$. The time-duration $\tau$ for $B$ to stay in the excited states $|g_{1},e_{2}\rangle$ or $|e_{1},g_{2}\rangle$ follows the probability $p(\tau)\propto e^{-\gamd_{1}\tau}$, and the system experiences multiple jumps 
\beq 
L^{\dagger}_{a}=\langle g_{1},e_{2}|e^{-i\tau H_{\rm int}} |e_{1},g_{2}\rangle=\sum_{n=0}^{\infty}\frac{(-ia)^{2n+1}}{(2n+1)!}  (L^{\dagger}L)^{n}L^{\dagger} \label{La}
\eeq
and
\beq
N_{a}=\langle e_{1},g_{2}|e^{-i\tau H_{\rm int}} |e_{1},g_{2}\rangle-1=\sum_{n=1}^{\infty}\frac{(-ia)^{2n}}{(2n)!}  (LL^{\dagger})^{n} \label{Na}
\eeq
before $B$ returns to the ground state. In the above equations, a dimensionless parameter, $a=\lambda\tau$, is introduced. 

We can interpret that $\gamu_{1}$ is the average rate of the Poisson noise and its noise strength is given by $a$, which varies according to the distribution $p(a)\propto e^{-(\gamd_{1}/\lambda)a}$ of the time-duration when the bath is excited. The effect of Poisson noise on the system is indicated by the jump operators $L^{\dagger}_{a}$ and $N_{a}$. 

Similarly, when the second spin gets excited according to the rate $\gamu_{2}$, the system experiences multiple jumps $L_{a}$ and 
\beq
M_{a}=\langle g_{1},e_{2}|e^{-i\tau H_{\rm int}} |g_{1},e_{2}\rangle-1=\sum_{n=1}^{\infty}\frac{(-ia)^{2n}}{(2n)!}  (L^{\dagger}L)^{n} . \label{Ma}
\eeq
The above expressions indicate that infinite orders of the system-bath coupling strength and all $n$-point bath correlation functions should be considered. 

{\it $n$-point bath correlation function.---} We now calculate the $n$-point bath correlation functions. We introduce the notation $\rho_{\rm B}^{+}=|g_{1},e_{2}\rangle\langle g_{1},e_{2}|$ and $\rho_{\rm B}^{-}=|e_{1},g_{2}\rangle\langle e_{1},g_{2}|$. 
The $2n$-point correlation functions can be decomposed into a product form
\begin{widetext}
\beq
\chi_{2n}^{\vec{l},\vec{k}}(t_{1},\cdots,t_{2n})=\prod_{j=1}^{n-1}\Tr_{\rm B}[B^{k_{2j-1}}_{l_{2j-1}}e^{\mathcal{L}_{\rm B}(t_{2j-1}-t_{2j})}B^{k_{2j}}_{l_{2j}}e^{\mathcal{L}_{\rm B}(t_{2j}-t_{2j+1})}\mathcal{Q}\rho_{\rm B}^{l_{2j+1}k_{2j+1}}]  \Tr_{\rm B}[B^{k_{2n-1}}_{l_{2n-1}}e^{\mathcal{L}_{\rm B}(t_{2n-1}-t_{2n})}B^{k_{2n}}_{l_{2n}}\rho_{\rm B}^{\rm eq}], \label{generalchi}
\eeq
\end{widetext}
where all odd-point correlation functions vanish. In the supplemental material~\cite{supplement}, we give an explicit expression for each term appearing in Eq.~(\ref{generalchi}).  

%Based on the equivalence of the reduced dynamics of the system with the same bath correlation functions, various non-Gaussian baths may be described by combining a sufficient number of random telegraph noise baths. To this end, the analytical expression of the bath correlation functions~(\ref{generalchi}) may be utilized~\cite{supplement}. 
%Similar techniques for Gaussian baths are known, for example, as the pseudomode method~\cite{pseudo1, pseudo2, pseudo3, pseudo4}, reaction-coordinate mapping method~\cite{RC1, RC2, RC3}, and mesoscopic leads approach combined with tensor-network techniques~\cite{tensor}.

{\it Poisson white noise limit and the master equation.---} 
Poisson white noise limit is obtained by considering $\Gamma_{1}^{-}=\Gamma_{2}^{-} \rightarrow \infty$ and $\lambda\rightarrow \infty$ while fixing the effective coupling strength $\mu=\lambda/\Gamma_{1}^{-}$. Consequently, the $2n$-point bath correlation functions take the following form: 
\beq
\chi_{2n}^{\vec{l},\vec{k}} 
= 
\begin{cases}
\frac{1}{2} \gamu_{1}
(2\mu)^{2n} \prod_{j=1}^{2n-1} \delta(t_{j}-t_{j+1}) & \text{if } l_{2n}k_{2n}=+ 
\\ 
\frac{1}{2} \gamu_{2}
(2\mu)^{2n} \prod_{j=1}^{2n-1} \delta(t_{j}-t_{j+1}) & \text{if } l_{2n}k_{2n}=-
\end{cases}
\label{2nCPoisson} 
\eeq
where $\vec{l}$ and $\vec{k}$ satisfy the condition of $k_{2i-1}=-k_{2i}$  $(1\leq i\leq n)$ and $l_{2j+1}k_{2j+1}=-l_{2j}k_{2j}$ $(1\leq j\leq n-1)$; otherwise, $\chi_{2n}^{\vec{l},\vec{k}}$ vanishes. 

Substituting Eq.~(\ref{2nCPoisson}) into Eq.~(\ref{Seq}) yields the following master equation as the Poisson white noise limit of the bath, 
\beqa
\partial_{t}\rho_{\rm S} 
&=& 
-i[H_{\rm S},\rho_{\rm S}]
+ 
\int^{\infty}_{0}da \, p(a)  
\Bigl[ \Gamma^{+}_{2} ( \mathcal{D}_{L_{a}}[\rho_{\rm S}] + \mathcal{D}_{M_{a}}[\rho_{\rm S}] ) 
\nonumber \\
& &+ 
\Gamma^{+}_{1}( \mathcal{D}_{L_{a}^{\dagger}}[\rho_{\rm S}] + \mathcal{D}_{N_{a}}[\rho_{\rm S}]) \Bigr],
\label{PoissonME}
\eeqa
where $\gamu_{1}$ and $\gamu_{2}$ are the noise rates; $a$ is the noise strength that varies according to the probability $p(a)=\mu^{-1}e^{-a/\mu}$; $\mu$ is the effective coupling strength; and $L^{\dagger}_{a}, N_{a}, M_{a}$ are the jump operators defined by Eqs.~(\ref{La}), (\ref{Na}), and (\ref{Ma}), respectively. For details, see Section {\it Multiple jump effects} for their physical intuitions. 

For a Hermitian system coupling operator, i.e., $L=L^{\dagger}=X$, the obtained master equation~(\ref{PoissonME}) reads $\partial_{t}\rho_{\rm S}=-i[H_{\rm S},\rho_{\rm S}]+(\gamu_{1}+\gamu_{2})\int_{0}^{\infty}da\, p(a)(\sin aX \rho_{\rm S}\sin aX + \cos aX\rho_{\rm S}\cos aX-\rho_{\rm S})$, which is consistent with the classical noise master equation~\cite{PoissonME3}. 

The Gaussian noise limit considers the noise occurring continuously but weakly, i.e., $\gamu_{i}\rightarrow \infty$ and $\mu\rightarrow 0$, by fixing $\mu^{2}\gamu_{i}$. In this limit, Eq.~(\ref{PoissonME}) reads 
\beq
\partial_{t}\rho_{\rm S} = -i[H_{\rm S},\rho_{\rm S}] + 2\mu^{2}\gamu_{2}\mathcal{D}_{L}[\rho_{\rm S}] + 2\mu^{2}\gamu_{1}\mathcal{D}_{L^{\dagger}}[\rho_{\rm S}] , \label{GaussianME}
\eeq
which is consistent with the conventional weak-coupling GKSL master equation~\cite{Breuer}. 

Note that the Markov approximation requires that the bath correlation time $\tau_{\rm B}$ is much smaller than the system relaxation time $\tau_{\rm R}$~\cite{Breuer}. For the Gaussian noise master equation~(\ref{GaussianME}), we have $\tau_{\rm B}=1/\gamd_{1} \ll \tau_{\rm R}=1/\lambda$ because $\mu\ll 1$. For the Poisson noise master equation~(\ref{PoissonME}), $\tau_{\rm R}$ is given by the time-scale in which the Poisson noise occurs. Therefore, the condition $\tau_{\rm B}=1/\gamd_{1}\ll\tau_{\rm R}=1/\gamu_{i}$ is valid although $\lambda$ and $\gamd_{1}$ is generically of the same order.

\begin{figure}[t]
\begin{center}
\includegraphics[width=.4\textwidth]{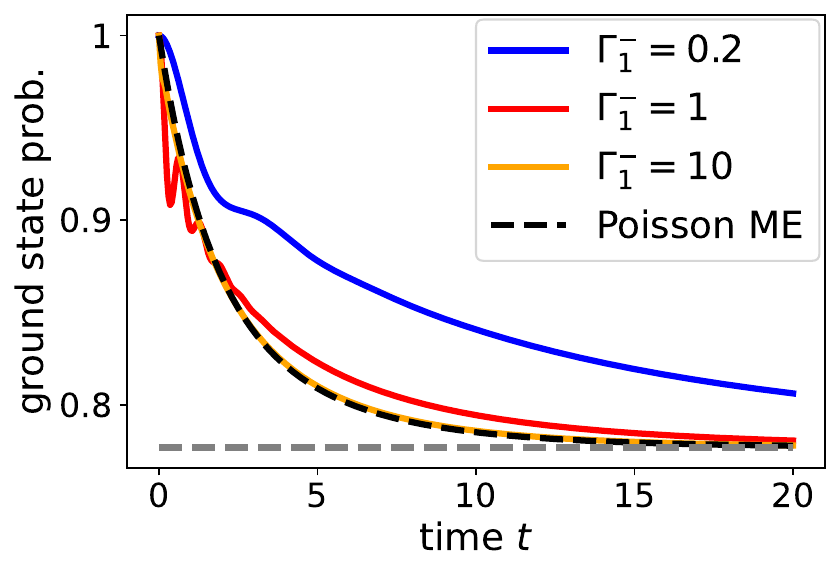}
\caption{
Plot of the ground state probability as a function of time for the collective system-bath coupling model. The black dotted curve is obtained from the Poisson noise master equation, Eq.~(\ref{PoissonME}). Dashed gray line is calculated by using the Gibbs state with inverse temperature $\beta$, i.e., $\rho_{\rm S}^{\rm eq}=e^{-\beta H_{\rm S}}/\Tr[e^{-\beta H_{\rm S}}]$. The solid curves are obtained by directly solving Eq.~(\ref{Eq1}) for different values of $\gamd_{1}=\gamd_{2}$ and $\lambda=\mu\gamd_{1}$. 
The figure shows that as $\gamd_{1}$ increases, the condition $\gamu_{i}\ll\gamd_{1},\lambda$ is better satisfied and converges to the result calculated by Eq.~(\ref{PoissonME}). 
%The figure also shows that the system relaxes to the thermal equilibrium state, thereby demonstrating that Eq.~(\ref{PoissonME}) satisfies the detailed balance condition for the model considered here. 
The initial state is given by the ground state, and the parameters are 
$N=6, \mu=2.0, \beta=1.5, \omega=\omega_{1}-\omega_{2}=1, \gamu_{2}=1$.
}
\label{fig_DB}
\end{center}
\end{figure}

\begin{figure}[t]
\begin{center}
\includegraphics[width=.4\textwidth]{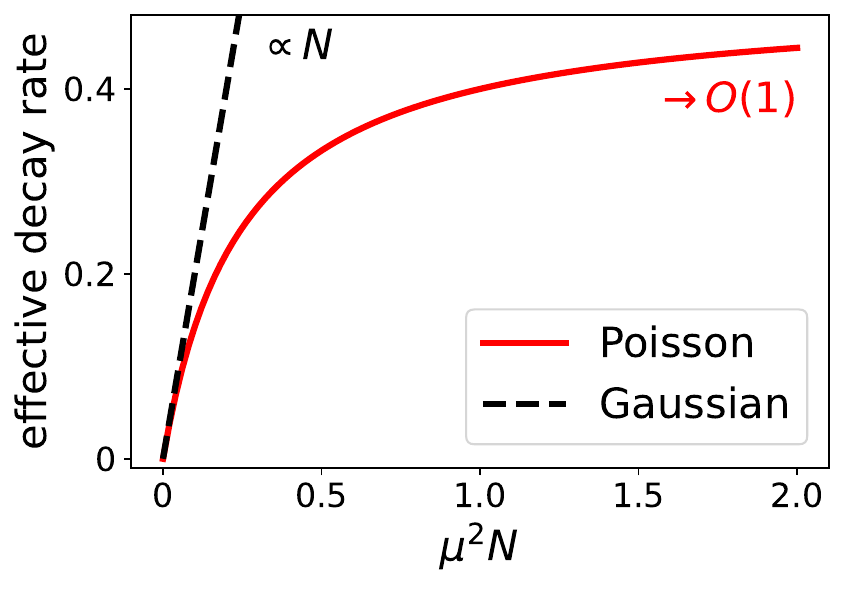}
\caption{ Plot of the effective decay rate $\Gamma_{\rm eff}/\gamu_{2}$ [Eq.~(\ref{effdecay})], as a function of $\mu^{2}N$, where $N$ is the number of two-level systems and $\mu$ is the effective coupling strength. The red curve is obtained using the Poisson noise master equation~(\ref{PoissonME}), which converges to the rate $\Gamma_{\rm eff}=\gamu_{2}/2=O(1)$ in large $N$, where $\gamu_{2}$ is the average rate of the Poisson noise. Gaussian noise limit corresponds to $\mu \ll 1$ and agrees with the $O(N)$ scaling of the effective decay rate for the conventional Gaussian noise master equation~(\ref{GaussianME}) analysis, plotted as the black dashed line.
}
\label{fig_sr}
\end{center}
\end{figure}

{\it Example: collective system-bath coupling.---} Suppose that the system is given by $N$ two-level systems with the Hamiltonian $H_{\rm S}=(\omega/2)\sum_{i}\sigma_{i}^{z}$. When the system collectively couples with the bath such that the system coupling operator is $L=\sum_{i}\sigma^{-}_{i}$, its effective decay rate is enhanced by a factor $N$, which is termed as super-radiance~\cite{Dicke}. However, the Poisson noise master equation has a different scaling for the effective decay rate. 

We first calculate the time-evolution of the system, which is plotted in Fig.~\ref{fig_DB}. This figure demonstrates that the Poisson noise master equation~(\ref{PoissonME}) is valid when $\gamu_{i}\ll\gamd_{1},\gamd_{2},\lambda$ while fixing $\mu=\lambda/\gamd_{1}$. By setting $\omega=\omega_{1}-\omega_{2}$, the rates satisfy $\gamu_{2}/\gamu_{1}=e^{\beta\omega}$, and the detailed balance condition is satisfied. See Fig.~\ref{fig_DB}~\cite{footnote}.  

Next, we assume $\Gamma^{+}_{1}=0$ for simplicity and discuss the spontaneous decay to the ground state.  
The effective decay rate $\Gamma_{\rm eff}$, or the transition rate, from the first excited Dicke state~\cite{Dicke} $\lvert D_{N,1}\rangle = (\lvert e,g,\cdots,g\rangle+\text{all permutations})/\sqrt{N}$ to the ground state $\lvert 0 \rangle = \lvert g,\cdots,g\rangle$ reads
\beq
\Gamma_{\rm eff}=\Gamma_{2}^{+}\int^{\infty}_{0}da \, p(a) |\langle 0|L_{a}|D_{N,1}\rangle|^{2}=\frac{2\Gamma^{+}_{2}\mu^{2}N }{1+4\mu^{2}N}. \label{effdecay}
\eeq
In the Gaussian noise limit $\mu \ll 1$, the effective decay rate is given by $2\Gamma^{+}_{2}\mu^{2}N$, which reproduces the conventional $O(N)$ enhancement of the decay rate. On the other hand, when $N$ or $\mu$ become sufficiently large, i.e., $\mu^{2}N\gg 1$, the effective decay rate converges to $\Gamma^{+}_{2}/2$, and the scaling is $O(1)$ for the Poisson noise. These distinct scaling behaviors are plotted in Fig.~\ref{fig_sr}, and their physical origins can be understood from the bath noise properties. The system decay time-scale cannot surpass the average rate $\gamu_{2}$ of Poisson noise. On the other hand, Gaussian noise occurs continuously, and its noise strength depends quadratically on the enhanced system-bath coupling $(=\mu^{2}N)$.

{\it Conclusions.---} We have derived a Markovian quantum master equation in which the non-Gaussian bath exhibits Poisson noise properties. We showed that the system experiences multiple nonlinear jumps in terms of the system-bath interaction Hamiltonian, whereas the jump operators are linear in the conventional weak-coupling master equation. For the application, we considered a collective coupling between the bath and $N$ two-level systems. We revealed that the effective decay rate saturates as $O(1)$ for the Poisson noise master equation, whereas it scales as $O(N)$ for the conventional case. This result demonstrates that the difference in the bath statistics drastically alters the dissipative properties of quantum systems. 

Our results not only develop a non-Gaussian bath theory in the white noise regime, but also provide an essential step towards establishing a general theory of open quantum dynamics under the influence of generic baths beyond Gaussian properties.
Further elucidation and characterization of the dissipation and decoherence induced by non-Gaussian baths are crucial 
\revision{to develop a formalism that can describe the effect of two-level defect baths on the system in solid-state quantum devices~\cite{1fnoise, twolevelbath1, twolevelbath2, twolevelbath3},} 
to design dissipative quantum state engineering~\cite{Cirac09, tunable4}, and to comprehend the transport properties in high-performance photovoltaic and optoelectronic devices. Understanding the role of non-Gaussian bath statistics is anticipated to lead to remarkable progress in such multidisciplinary research fields that span from quantum science to materials science. 

\revision{
As one of the challenging but important future directions, we point out that developing a method to calculate $n$-point bath correlation functions for various bath models allows one to study the influence of non-Gaussian baths to the system based on Eq.~(\ref{Seq}). Moreover, it is also challenging but worth further investigation to develop a numerical technique to efficiently calculate Eq.~(\ref{Seq}) by extending the techniques developed for Gaussian baths, for example, the pseudomode method~\cite{pseudo1, pseudo2, pseudo3, pseudo4}, reaction-coordinate mapping method~\cite{RC1, RC2, RC3}, and mesoscopic leads approach combined with tensor-network techniques~\cite{tensor}. 
%Based on the equivalence of the reduced dynamics of the system with the same bath correlation functions, various non-Gaussian baths may be described by combining a sufficient number of random telegraph noise baths. To this end, the analytical expression of the bath correlation functions~(\ref{generalchi}) may be utilized~\cite{supplement}. 
%Similar techniques for Gaussian baths are known, for example, as the pseudomode method~\cite{pseudo1, pseudo2, pseudo3, pseudo4}, reaction-coordinate mapping method~\cite{RC1, RC2, RC3}, and mesoscopic leads approach combined with tensor-network techniques~\cite{tensor}.
}

{\it Acknowledgements.---}
\revision{
The numerical calculations were done by using the QuTiP library~\cite{Qutip1,Qutip2}. 
We thank Kiyoshi Kanazawa for useful discussions.} 
K.F. acknowledges support from JSPS KAKENHI (Grant Number 23K13036).
This work was supported by MEXT Quantum Leap Flagship Program (Grant Number JPMXS0120330644) and JSPS KAKENHI (Grant Number JP21H01052).

\appendix

\widetext

\section{Notations}
The time-evolution equation of the bath $B$ is given by
\beq
\partial_{t}\rho_{\rm B}=\calL_{\rm B}\rho_{\rm B}= \sum_{j=1,2}\calL_{\rm B}^{j}\rho_{\rm B} \label{BathME}
\eeq
and
\beq
\calL_{\rm B}^{j} =-i[H^{j}_{\rm B}, \bullet]+ \Gamma^{+}_{j}\mathcal{D}_{\sigma^{+}_{j}}[\bullet]+\gamd_{j}\mathcal{D}_{\sigma^{-}_{j}}[\bullet] , \label{LE1}
\eeq
where $H_{\rm B}^{j}=(\omega_{j}/2)\sigma^{z}_{j}$ is the Hamiltonian of the $j$-th spin and $\Gamma_{j}^{\pm}$ are the jump rates. 
To simplify the notation, we introduce $\gamma_{j}=\Gamma^{+}_{j}+\Gamma^{-}_{j}$. The equilibrium steady-state is given by
\beq
\rho_{\rm B}^{\rm eq} = \rho_{\rm B}^{ {\rm eq},1} \otimes  \rho_{\rm B}^{ {\rm eq},2} \label{rhoBeq}
\eeq
where
\beq
\rho_{\rm B}^{ {\rm eq},j} =\frac{1}{\gamma_{j}}  \Bigr( \gamd_{j}|g\rangle\langle g|_{j} + \gamu_{j} |e\rangle \langle e|_{j} \Bigr) .
\eeq

In what follows, we calculate the $2n$-point bath correlation functions of the bath coupling super-operator $B^{k}_{l}$:
\beq
\chi^{ \vec{l},\vec{k} }_{n}(t_{1},\cdots,t_{n})
= 
\Tr
\left[ 
B^{k_{1}}_{l_{1}}
e^{\mathcal{L}_{\rm B}(t_{1}-t_{2})}
\calQ 
B^{k_{2}}_{l_{2}} 
e^{\mathcal{L}_{\rm B}(t_{2}-t_{3})}
\cdots 
e^{\mathcal{L}_{\rm B}(t_{n-1}-t_{n}) } 
\calQ 
B^{k_{n}}_{l_{n}} 
\rho_{\rm B}^{\rm eq}
\right] , 
\label{sup:bathC}
\eeq 
where $B^{\pm}_{+}\bullet =(\lambda\sigma_{1}^{\mp}\sigma_{2}^{\pm})\bullet$ and $B^{\pm}_{-}\bullet = \bullet (\lambda\sigma_{1}^{\mp}\sigma_{2}^{\pm})$.

\section{$2n$-point bath correlation functions}
When calculating the $2n$-point bath correlation functions, we utilize the following relations:
\beqa
e^{\mathcal{L}_{\rm B}^{j}t}\sigma_{j}^{+} &=& e^{-\gamma_{j} t/2}e^{-i\omega_{j} t}\sigma_{j}^{+} \nonumber \\
e^{\mathcal{L}_{\rm B}^{j}t}\sigma_{j}^{-} &=& e^{-\gamma_{j} t/2}e^{i\omega_{j} t}\sigma_{j}^{-} \nonumber \\
e^{\mathcal{L}_{\rm B}^{j}t}\sigma_{j}^{z} &=& e^{-\gamma_{j} t}\sigma_{j}^{z}  \nonumber \\
e^{\mathcal{L}_{\rm B}^{j}t} |e\rangle\langle e|_{j} &=& \rho_{\rm B}^{ {\rm eq},j} + \frac{\gamd_{j} }{\gamma_{j}}e^{-\gamma_{j} t} \sigma_{j}^{z} \nonumber \\
e^{\mathcal{L}_{\rm B}^{j}t} |g\rangle\langle g|_{j} &=& \rho_{\rm B}^{ {\rm eq},j} - \frac{\gamu_{j}}{\gamma_{j}}e^{-\gamma_{j} t}\sigma_{j}^{z}. \label{relation}
\eeqa

\subsection{Two-point bath correlation functions}
We first note that $\Tr[B^{k}_{l}\rho_{\rm B}^{\rm eq}]=0$, and therefore $\Tr[B_{l_{1}}^{k_{1}}e^{\mathcal{L}t} \mathcal{Q} B^{k_{2}}_{l_{2}}\rho_{\rm B}^{\rm eq}]=\Tr[B_{l_{1}}^{k_{1}}e^{\mathcal{L}t}  B^{k_{2}}_{l_{2}}\rho_{\rm B}^{\rm eq}]$. 
Using Eq.~(\ref{relation}), the two-point bath correlation functions can be calculated and they are given by
\beqa
 \Tr[B_{l}^{-}e^{\mathcal{L}_{\rm B}t}B^{+}_{+}\rho_{\rm B}^{\rm eq}] &=& \frac{\lambda^{2}\gamu_{1}\gamd_{2}}{ \gamma_{1}\gamma_{2}}e^{i(\omega_{1}-\omega_{2})t}e^{-\frac{1}{2}(\gamma_{1}+\gamma_{2})t}, \nonumber \\
\Tr[B^{+}_{l}e^{\mathcal{L}_{\rm B}t}B^{-}_{-}\rho_{\rm B}^{\rm eq}] &=& \frac{\lambda^{2}\gamu_{1}\gamd_{2}}{ \gamma_{1}\gamma_{2}}e^{-i(\omega_{1}-\omega_{2})t}e^{-\frac{1}{2}(\gamma_{1}+\gamma_{2})t}, \nonumber \\
\Tr[B_{l}^{-}e^{\mathcal{L}_{\rm B}t}B^{+}_{-}\rho_{\rm B}^{\rm eq}] &=& \frac{\lambda^{2}\gamd_{1}\gamu_{2}}{ \gamma_{1}\gamma_{2}}e^{i(\omega_{1}-\omega_{2})t}e^{-\frac{1}{2}(\gamma_{1}+\gamma_{2})t}, \nonumber \\
\Tr[B^{+}_{l}e^{\mathcal{L}_{\rm B}t}B^{-}_{+}\rho_{\rm B}^{\rm eq}] &=& \frac{\lambda^{2}\gamd_{1}\gamu_{2}}{ \gamma_{1}\gamma_{2}}e^{-i(\omega_{1}-\omega_{2})t}e^{-\frac{1}{2}(\gamma_{1}+\gamma_{2})t} . \label{twopoint}
\eeqa
Other combinations vanish, i.e., $\Tr[B_{l_{1}}^{k_{1}}e^{\mathcal{L}_{\rm B}t}B^{k_{2}}_{l_{2}}\rho_{\rm B}^{\rm eq}]=0$ if $k_{1}=k_{2}$.

\subsection{Two-point bath correlation function-like quantities}
The two-point bath correlation function-like quantities starting from the initial state $\rho_{\rm B}^{-}=|e_{1},g_{2}\rangle\langle e_{1},g_{2}|$ are defined as
\beqa
\Tr[B^{-}_{l}e^{\mathcal{L}_{\rm B}t}B^{+}_{+}e^{\mathcal{L}_{\rm B}s}\calQ\rho_{\rm B}^{-}]&=&\frac{\lambda^{2} }{ \gamma_{1}\gamma_{2} } \Bigl[(\Gamma_{1}^{-}e^{-\gamma_{1}s}+\gamu_{1})(\gamd_{2}+\gamu_{2}e^{-\gamma_{2}s} )-\gamu_{1}\gamd_{2} \Bigr] e^{i(\omega_{1}-\omega_{2})t }e^{-\frac{1}{2}(\gamma_{1}+\gamma_{2})t} , \nonumber \\
\Tr[B^{+}_{l}e^{\mathcal{L}_{\rm B}t}B^{-}_{-}e^{\mathcal{L}_{\rm B}s}\calQ\rho_{\rm B}^{-}]&=&\frac{\lambda^{2} }{ \gamma_{1}\gamma_{2} } \Bigl[(\Gamma_{1}^{-}e^{-\gamma_{1}s}+\gamu_{1})(\gamd_{2}+\gamu_{2}e^{-\gamma_{2}s} )-\gamu_{1}\gamd_{2} \Bigr] e^{-i(\omega_{1}-\omega_{2})t }e^{-\frac{1}{2}(\gamma_{1}+\gamma_{2})t}, \nonumber \\
\Tr[B^{-}_{l}e^{\mathcal{L}_{\rm B}t}B^{+}_{-}e^{\mathcal{L}_{\rm B}s}\calQ \rho_{\rm B}^{-}] &=& \frac{\lambda^{2}\Gamma_{1}^{-}\Gamma_{2}^{+}}{\gamma_{1}\gamma_{2}} \Bigl[ (1-e^{-\gamma_{1}s })(1- e^{-\gamma_{2}s }) -1 \Bigr] e^{i(\omega_{1}-\omega_{2})t }e^{-\frac{1}{2}(\gamma_{1}+\gamma_{2})t} , \nonumber \\
\Tr[B^{+}_{l}e^{\mathcal{L}_{\rm B}t}B^{-}_{+}e^{\mathcal{L}_{\rm B}s} \calQ \rho_{\rm B}^{-}] &=& \frac{\lambda^{2}\Gamma_{1}^{-}\Gamma_{2}^{+}}{\gamma_{1}\gamma_{2}} \Bigl[ (1-e^{-\gamma_{1}s })(1- e^{-\gamma_{2}s }) -1\Bigr] e^{-i(\omega_{1}-\omega_{2})t }e^{-\frac{1}{2}(\gamma_{1}+\gamma_{2})t} . \label{twopointeg}
\eeqa
Other combinations vanish, i.e., $\Tr[B^{k_{1}}_{l_{1}}e^{\mathcal{L}_{\rm B}t}B^{k_{2}}_{l_{2}}e^{\mathcal{L}_{\rm B}s}\calQ\rho_{\rm B}^{-}]=0$ if $k_{1}=k_{2}$. 

Similarly, two-point correlation function-like quantities starting from the initial state $\rho_{\rm B}^{+}=|g_{1},e_{2}\rangle\langle g_{1},e_{2}|$ are defined as:
\beqa
\Tr[B^{+}_{l}e^{\mathcal{L}_{\rm B}t}B^{-}_{+}e^{\mathcal{L}_{\rm B}s}\calQ \rho_{\rm B}^{+}]&=&\frac{\lambda^{2} }{ \gamma_{1}\gamma_{2} } \Bigl[(\Gamma_{2}^{-}e^{-\gamma_{1}s}+\Gamma_{2}^{+})(\Gamma_{1}^{-}+\Gamma_{1}^{+}e^{-\gamma_{2}s})-\gamu_{2}\gamd_{1}\Bigr]e^{-i(\omega_{1}-\omega_{2})t }e^{-\frac{1}{2}(\gamma_{1}+\gamma_{2})t}, \nonumber  \\
\Tr[B^{-}_{l}e^{\mathcal{L}_{\rm B}t}B^{+}_{-}e^{\mathcal{L}_{\rm B}s}\calQ \rho_{\rm B}^{+}]&=&\frac{\lambda^{2} }{ \gamma_{1}\gamma_{2} } \Bigl[(\Gamma_{2}^{-}e^{-\gamma_{1}s}+\Gamma_{2}^{+})(\Gamma_{1}^{-}+\Gamma_{1}^{+}e^{-\gamma_{2}s})-\gamu_{2}\gamd_{1}\Bigr] e^{i(\omega_{1}-\omega_{2})t }e^{-\frac{1}{2}(\gamma_{1}+\gamma_{2})t} \nonumber \\
\Tr[B^{+}_{l}e^{\mathcal{L}_{\rm B}t}B^{-}_{-}e^{\mathcal{L}_{\rm B}s}\calQ \rho_{\rm B}^{+}] &=& \frac{\lambda^{2}\Gamma_{1}^{+}\Gamma_{2}^{-}}{\gamma_{1}\gamma_{2}} \Bigl[ (1-e^{-\gamma_{1}s })(1- e^{-\gamma_{2}s }) -1\Bigr] e^{-i(\omega_{1}-\omega_{2})t }e^{-\frac{1}{2}(\gamma_{1}+\gamma_{2})t} \nonumber \\
\Tr[B^{-}_{l}e^{\mathcal{L}_{\rm B}t}B^{+}_{+}e^{\mathcal{L}_{\rm B}s}\calQ \rho_{\rm B}^{+}] &=& \frac{\lambda^{2}\Gamma_{1}^{-}\Gamma_{2}^{+}}{\gamma_{1}\gamma_{2}} \Bigl[ (1-e^{-\gamma_{1}s })(1- e^{-\gamma_{2}s }) -1\Bigr] e^{i(\omega_{1}-\omega_{2})t }e^{-\frac{1}{2}(\gamma_{1}+\gamma_{2})t} . \label{twopointge}
\eeqa
Other combinations vanish, i.e., $\Tr[B^{k_{1}}_{l_{1}}e^{\mathcal{L}_{\rm B}t}B^{k_{2}}_{l_{2}}e^{\mathcal{L}_{\rm B}s}\calQ\rho_{\rm B}^{+}]=0$ if $k_{1}=k_{2}$. 

\subsection{Decomposition of the $2n$-point bath correlation function into a product form}
We now focus on the following properties:
\beqa
B_{l_{1}}^{k_{1}}e^{\mathcal{L}_{\rm B}t}\calQ B^{k_{2}}_{l_{2}}\rho_{\rm B}^{\rm eq} &=& \begin{cases}
\rho_{\rm B}^{-} \Tr[B_{l_{1}}^{k_{1}}e^{\mathcal{L}_{\rm B}t}B^{k_{2}}_{l_{2}}\rho_{\rm B}^{\rm eq} ] & \text{if } l_{1}k_{1}=-, 
\\ 
\rho_{\rm B}^{+} \Tr[B_{l_{1}}^{k_{1}}e^{\mathcal{L}_{\rm B}t}B^{k_{2}}_{l_{2}}\rho_{\rm B}^{\rm eq} ] & \text{if } l_{1}k_{1}=+.  
\end{cases} \nonumber \\
B^{k_{1}}_{l_{1}}e^{\mathcal{L}_{\rm B}t} \calQ B^{k_{2}}_{l_{2}}e^{\mathcal{L}_{\rm B}s}\calQ\rho_{\rm B}^{\pm} &=& 
\begin{cases}
\rho_{\rm B}^{-} \Tr[B^{k_{1}}_{l_{1}}e^{\mathcal{L}_{\rm B}t}B^{k_{2}}_{l_{2}}e^{\mathcal{L}_{\rm B}s}\calQ\rho_{\rm B}^{\pm} ] & \text{if } l_{1}k_{1}=-, 
\\ 
\rho_{\rm B}^{+} \Tr[B^{k_{1}}_{l_{1}}e^{\mathcal{L}_{\rm B}t}B^{k_{2}}_{l_{2}}e^{\mathcal{L}_{\rm B}s}\calQ\rho_{\rm B}^{\pm}] & \text{if } l_{1}k_{1}=+.  
\end{cases}
\eeqa
The above relations allows us to decompose the $2n$-point bath correlation functions into a product form
\beqa
\chi_{2n}^{\vec{l},\vec{k}}(t_{1},\cdots,t_{2n})&=&\prod_{j=1}^{n-1}\Tr[B^{k_{2j-1}}_{l_{2j-1}}e^{\mathcal{L}_{\rm B}(t_{2j-1}-t_{2j})}B^{k_{2j}}_{l_{2j}}e^{\mathcal{L}_{\rm B}(t_{2j}-t_{2j+1})}\mathcal{Q}\rho_{\rm B}^{l_{2j+1}k_{2j+1}}] \nonumber \\
& &\times  \Tr[B^{k_{2n-1}}_{l_{2n-1}}e^{\mathcal{L}_{\rm B}(t_{2n-1}-t_{2n})}B^{k_{2n}}_{l_{2n}}\rho_{\rm B}^{\rm eq}]. \label{2nchi}
\eeqa
Note that odd-point bath correlation functions vanish because $\Tr[B^{k_{2}}_{l_{2}}e^{\mathcal{L}_{\rm B}s}\calQ\rho_{\rm B}^{\pm}]=0$. 

\subsection{Poisson white noise limit}
We next consider the Poisson white noise by taking the limit $\Gamma_{1}^{-}=\Gamma_{2}^{-} \rightarrow \infty$ and $\lambda\rightarrow \infty$ while fixing $\lambda/\Gamma_{1}^{-}=\mu$. We use the delta function formula $a e^{-a |t|}\rightarrow 2\delta(t)$ for $a\rightarrow \infty$. 

The two-point correlation function reads
\beq
\Tr[B^{k_{2n-1}}_{l_{2n-1}}e^{\mathcal{L}_{\rm B}(t_{2n-1}-t_{2n})}B^{k_{2n}}_{l_{2n}}\rho_{\rm eq}] \rightarrow 
\begin{cases}
 2\mu^{2}\gamu_{1} \delta_{k_{2n-1},-k_{2n}}\delta(t_{2n-1}-t_{2n}) & \text{if } l_{2n}k_{2n}=1 \\
 2\mu^{2}\gamu_{2} \delta_{k_{2n-1},-k_{2n}}\delta(t_{2n-1}-t_{2n}) & \text{if } l_{2n}k_{2n}=-1 .
\end{cases}
\eeq
Similarly, we have
\beqa
& &\Tr[B^{k_{2j-1}}_{l_{2j-1}}e^{\mathcal{L}_{\rm B}(t_{2j-1}-t_{2j})}B^{k_{2j}}_{l_{2j}}e^{\mathcal{L}_{\rm B}(t_{2j}-t_{2j+1})}\mathcal{Q}\rho_{\rm B}^{l_{2j+1}k_{2j+1}}] \\
& &\rightarrow \delta_{k_{2j-1},-k_{2j}}\delta_{l_{2j}k_{2j},-l_{2j+1}k_{2j+1}}4\mu^{2}\delta(t_{2j-1}-t_{2j})\delta(t_{2j}-t_{2j+1}).
\eeqa
Combining the above equations give the Poisson white noise limit of the $2n$-point bath correlation functions: 
\beq
\chi_{2n}^{\vec{l},\vec{k}} (t_{1},\cdots,t_{2n})
= 
\begin{cases}
\frac{1}{2} \gamu_{1}
(2\mu)^{2n} \prod_{j=1}^{2n-1} \delta(t_{j}-t_{j+1}) & \text{if } l_{2n}k_{2n}=+, 
\\ 
\frac{1}{2} \gamu_{2}
(2\mu)^{2n} \prod_{j=1}^{2n-1} \delta(t_{j}-t_{j+1}) & \text{if } l_{2n}k_{2n}=-. 
\end{cases}
\label{sup:2nCPoisson} 
\eeq
with the constraint $k_{2i-1}=-k_{2i}$  $(1\leq i\leq n)$ and $l_{2j+1}k_{2j+1}=-l_{2j}k_{2j}$ $(1\leq j\leq n-1)$. 

By using Eq.~(\ref{2nCPoisson}), the master equation reads
\beqa
\partial_{t}\rho_{S}(t)&=&-i[H_{S},\rho_{S}] + \Gamma_{2}^{+} \sum_{n=1}^{\infty}\sum_{j=0}^{2n} \mu^{2n}(-1)^{j+n}{}_{2n}C_{j}  \overbrace{\cdots LL^{\dagger}L}^{j}\rho_{S} \overbrace{ L^{\dagger}LL^{\dagger}\cdots}^{2n-j}  \nonumber \\
& &+ \Gamma_{1}^{+} \sum_{n=1}^{\infty}\sum_{j=0}^{2n} \mu^{2n}(-1)^{j+n}{}_{2n}C_{j}  \overbrace{\cdots L^{\dagger}LL^{\dagger}}^{j}\rho_{S} \overbrace{ LL^{\dagger}L\cdots}^{2n-j}  .
\eeqa
Further transforming the above expression into the GKSL form gives 
\beq
\partial_{t}\rho_{\rm S} 
= \mathcal{L}^{\rm Poisson}\rho_{\rm S} =
-i[H_{\rm S},\rho_{\rm S}]
+ 
\int^{\infty}_{0}da \, p(a)  
\Bigl[ \Gamma^{+}_{2} ( \mathcal{D}_{L_{a}}[\rho_{\rm S}] + \mathcal{D}_{M_{a}}[\rho_{\rm S}] ) + 
\Gamma^{+}_{1}( \mathcal{D}_{L_{a}^{\dagger}}[\rho_{\rm S}] + \mathcal{D}_{N_{a}}[\rho_{\rm S}]) \Bigr],
\label{sup:PoissonME}
\eeq
presented in the main text. 

\revision{
\section{Simulating non-Gaussian baths by combining random telegraph noise baths}
Based on the equivalence of the reduced dynamics of the system with the same bath correlation functions, the analytical expression of the bath correlation functions~(\ref{twopoint}), (\ref{twopointeg}), (\ref{twopointge}) and (\ref{2nchi}) may be utilized to model various non-Gaussian baths. To this end, we decompose the bath correlation functions of a given non-Gaussian bath into a finite sum of exponentially decaying terms in time as $\chi_{2n}^{\vec{l},\vec{k}}=\sum_{m=1}^{M} f^{(m)}(\vec{l},\vec{k},t_{1},\cdots,t_{2n})$, where
\beq
f^{(m)}(\vec{l},\vec{k},t_{1},\cdots,t_{2n}) = \sum_{m=1}^{M} a^{(m)}\prod_{j=1}^{2n-1} e^{ -\nu_{j}^{(m)}(t_{j}-t_{j+1}) }.
\eeq
Here, $a^{(m)}$ and $\nu_{j}^{(m)}$ are some parameters. 
Then, we fit each exponential terms with Eq.~(\ref{2nchi}). It should be noted that fitting not only the two-time bath correlation functions, but also higher-order terms, requires huge complexity. Moreover, the flexibility of designing the functional form of the bath correlation functions by using the random telegraph noise bath model is quite limited. To overcome these difficulties, we address the the techniques developed for Gaussian baths, especially the pseudomode method~\cite{pseudo1, pseudo2, pseudo3, pseudo4}, which extends the bath parameters into complex-valued quantities, allowing more flexibility to describe a wide range of Gaussian baths. Extending the pseudomode methods to non-Gaussian baths by using (complex parameter-valued) damped few-spin systems is left for future work.

\section{Comparison between the exact system-bath dynamics and the Poisson noise master equation}
In this section, we give additional information about how the time-evolution of the system described by the exact system-bath dynamics 
\beq
\partial_{t}\rho_{\rm SB}(t) =  -i[H_{\rm S}+H_{\rm int},\rho_{\rm SB}] + \sum_{j=1,2}\calL^{j}_{\rm B} \rho_{\rm SB} , 
\label{sup:Eq1}
\eeq
converges to that described by the Poisson noise master equation~(\ref{PoissonME}), where $H_{\rm int}=\lambda(L\sigma_{1}^{+}\sigma_{2}^{-}+L^{\dagger}\sigma_{1}^{-}\sigma_{2}^{+})$ and $\calL^{j}_{\rm B}$ is defined in Eq.~(\ref{LE1}). We consider the collective system-bath coupling model 
\beq
H_{S}=(\omega/2)\sum_{i}\sigma_{i}^{z} \label{HS}
\eeq
and
\beq
L =\sum_{i}\sigma_{i}^{-}, \label{Lindblad}
\eeq
discussed in the main text. We assume that the initial state of the system is given by a superposition between the ground and the first excited state:
\beq
|\psi_{\rm S}(0)\rangle = \frac{1}{\sqrt{2}} ( |0\rangle + |D_{N,1}\rangle ).
\eeq
In Fig.~\ref{fig_decay}, we plot the ground state probability and the expectation value of $J_{x}=\sum_{i}\sigma_{i}^{x}$ as a function of time. The figure shows that as $\gamd_{1}$ increases, the Markovian condition $\gamu_{2}\ll\gamd_{1}$ is better satisfied and converges to the results calculated by using Eq.~(\ref{PoissonME}).

\begin{figure}[t]
\begin{center}
\includegraphics[width=.95\textwidth]{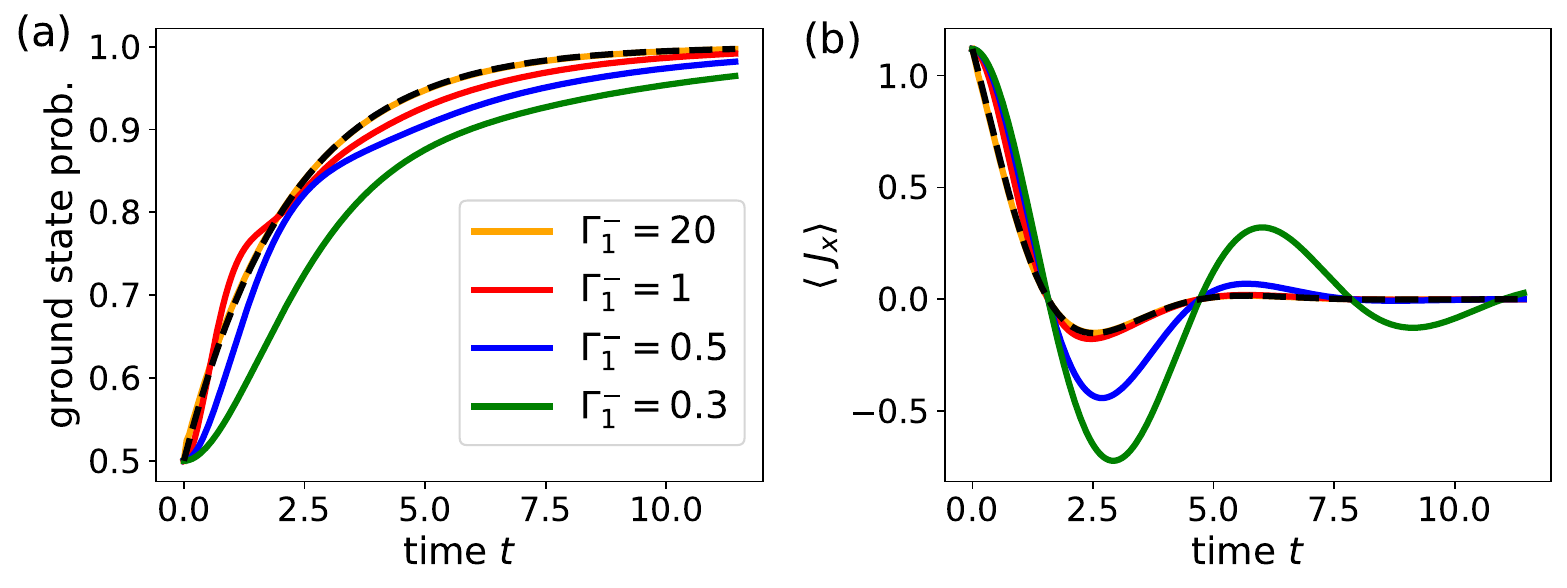}
\caption{\revision{Numerical calculation of the system-bath dynamics~(\ref{Eq1}) and the Poisson noise master equation~(\ref{PoissonME}) for the collective system-bath coupling model. 
(a) Plot of the ground state probability $\langle 0|\rho_{\rm S}|0\rangle$ as a function of time. (b) Plot of the expectation value of $J_x=\sum_{i}\sigma_{i}^{x}$ as a function of time. The initial state is given by $(|0\rangle + |D_{N,1}\rangle)/\sqrt{2}$. The black dotted curve is obtained from the Poisson noise master equation~(\ref{PoissonME}), and the solid curves are obtained by directly solving Eq.~(\ref{Eq1}) for different values of $\gamd_{1}=\gamd_{2}$ and $\lambda=\mu\gamd_{1}$. 
The figure shows that as $\gamd_{1}$ increases, the Markovian condition $\gamu_{2}\ll\gamd_{1}$ is better satisfied and converges to the results calculated by using Eq.~(\ref{PoissonME}). 
The parameters are $N=5, \mu=0.7, \gamu_{1}=0,\gamu_{2}=1,w=1, w_{1}=3,w_{2}=2$.}
}
\label{fig_decay}
\end{center}
\end{figure}

\section{Nakajima-Zwanzig projection operator method including the source field}
So far, we have shown that the reduced density matrix of the system $\rho_{\rm S}(t)$ follows Eq.~(\ref{PoissonME}) in the Poisson white noise limit of the bath. However, it is not obvious whether multi-time correlation functions of some system operators can be calculated with Eq.~(\ref{PoissonME}) in the same limit. To address this issue, we slightly generalize the Nakajima-Zwanzig projection operator method presented in the main text by including the source field of a system operator $A_{\rm S}$, which allows calculating the multi-time correlation functions of $A_{\rm S}$. We therefore modify the Liouville superoperator $\calL_{\rm S}$ as 
\beq
\calL_{\rm S}^{J}(t)\bullet =-i[H_{\rm S},\bullet ] -i J(t)A_{\rm S}\bullet,
\eeq
where $J(t)$ is the counting field. We then consider the following time-evolution equation 
\beq
\partial_{t}\chi^{J}_{\rm SB}(t) = \mathcal{L}^{J}(t) \chi^{J}_{\rm SB}(t), 
\label{EqJ}
\eeq
where $\mathcal{L}^{J}(t)=\calL_{\rm S}^{J}(t)+\calL_{\rm B}+\calL_{\rm int}$. Note that $\mathcal{L}_{\rm int}\bullet=-i[H_{\rm int},\bullet]$ and $\mathcal{L}_{\rm B}\bullet=-i[H_{\rm B},\bullet]+\mathcal{D}[\bullet]$. Also note that taking $J=0$ reproduces the Liouville superoperator used in the main text, i.e., $\calL^{J=0}(t)=\calL$. 

From Eq.~(\ref{EqJ}), we obtain the following formula for the multi-time correlation functions ($\tau\geq t_{1}\geq t_{2}\geq \cdots \geq t_{n}\geq 0$):
\beq
\langle A_{\rm S}(t_{1}) A_{\rm S}(t_{2}) \cdots A_{\rm S}(t_{n}) \rangle = i^{n}\frac{\delta^{n}}{\delta J(t_{1})\delta J(t_{2})\cdots \delta J(t_{n})}  \left. \Tr_{\rm SB}[ \chi^{J}_{\rm SB}(\tau) ] \right|_{J=0}, \label{multi-time}
\eeq
where 
\beq
\langle A_{\rm S}(t_{1}) A_{\rm S}(t_{2}) \cdots A_{\rm S}(t_{n}) \rangle = \Tr_{\rm SB}[A_{\rm S}e^{\calL(t_{1}-t_{2})}A_{\rm S} e^{\calL(t_{2}-t_{3})}A_{\rm S} \cdots A_{\rm S} e^{\calL(t_{n-1}-t_{n})} A_{\rm S} e^{\calL t_{n}} \rho_{\rm S}(0)\otimes \rho_{\rm B}^{\rm eq} ].
\eeq

\subsection{Nakajima-Zwanzig method}
In what follows, we discuss how the time-evolution equation for $\rho_{\rm S}(t)$ presented in the main text is modified by including the time-dependent counting field. First, Eq.~(2) in the main text is modified as
\beq
\partial_{t} \chi_{\rm S}^{J}(t)  =  \calL_{\rm S}^{J}(t) \chi_{\rm S}^{J}(t) + \int^{t}_{0} dv  \Tr_{\rm B} [ \calL_{\rm int} Te^{\int^{t}_{v}dx \mathcal{Q}\mathcal{L}^{J}(x) \mathcal{Q} } \mathcal{Q} \mathcal{L}_{\rm int} \rho_{\rm B}^{\rm eq}] \chi^{J}_{S}(v) \label{EqP2},
\eeq
where $\chi_{\rm S}^{J}(t)=\Tr_{\rm B}[\chi_{\rm SB}^{J}(t)]$ and $T$ is the time-ordering operator. We then use the following identity
\beq
Te^{\int^{t}_{v}dx \mathcal{Q}\mathcal{L}^{J}(x) \mathcal{Q} } = Te^{\int^{t}_{v}dx (\mathcal{L}^{J}_{\rm S}(x)+\calL_{\rm B})  } + \int^{t}_{v}dv_{1} Te^{\int^{t}_{v_{1}}dx (\mathcal{L}^{J}_{\rm S}(x)+\calL_{\rm B})  } \calQ \calL_{\rm int}\calQ Te^{\int^{v_{1}}_{v}dx_{1} \mathcal{Q}\mathcal{L}^{J}(x_{1}) \mathcal{Q}  } .\label{Identity}
\eeq
By expanding Eq.~(\ref{EqP2}) with the help of Eq.~(\ref{Identity}), we obtain
\beqa
\partial_{t} \chi_{\rm S}^{J}(t)  &=&  \calL_{\rm S}^{J}(t) \chi_{\rm S}^{J}(t) + \sum_{n=1}^{\infty} \int^{t}_{0}dv \int^{t}_{v}dv_{1}\int^{v_{1}}_{v}dv_{2}\cdots \int^{v_{n-2}}_{v}dv_{n-1} \nonumber \\
& &\times \Tr_{\rm B}[\calL_{\rm int}  Te^{\int^{t}_{v_{1}}dx_{1} (\mathcal{L}^{J}_{\rm S}(x_{1})+\calL_{\rm B})  } \calQ \calL_{\rm int}\calQ  Te^{\int^{v_{1}}_{v_{2}}dx_{2} (\mathcal{L}^{J}_{\rm S}(x_{2})+\calL_{\rm B})  } \nonumber \\
& &\hspace{10mm}  \cdots \calQ\calL_{\rm int}\calQ  Te^{\int^{v_{n-1}}_{v}dx_{n} (\mathcal{L}^{J}_{\rm S}(x_{n})+\calL_{\rm B})  } \calQ\calL_{\rm int}\rho_{\rm B}^{\rm eq} ] \chi_{\rm S}^{J}(v).
\eeqa
We then introduce a short-hand notation
\beq
\Lambda^{J}_{t,s}=Te^{\int^{t}_{s} dx \mathcal{L}^{J}_{\rm S}(x) },
\eeq
and write $Te^{\int^{t}_{v}dx (\mathcal{L}^{J}_{\rm S}(x)+\calL_{\rm B}) } = \Lambda^{J}_{t,v} e^{\calL_{\rm B}(t-v)} $ , since $\calL_{\rm S}^{J}(t)$ and $\calL_{\rm B}$ commute with each other. Using the above notations, we have
\beqa
\partial_{t} \chi_{\rm S}^{J}(t) &=& \calL_{\rm S}^{J}(t) \chi_{\rm S}^{J}(t) + \sum_{n=1}^{\infty} 
\int^{t}_{0}du_{1} \cdots \int^{u_{n-1}}_{0} du_{n} \sum_{\vec{l},\vec{k}} \prod_{j=1}^{n+1}(-i)^{l_{j}} \chi_{n+1}^{\vec{l},\vec{k}}(u_{1},u_{2},\cdots,u_{n},0) 
\nonumber \\
& &\times S^{k_{1}}_{l_{1}} \Lambda^{J}_{t,t-u_{1}+u_{2}} S^{k_{2}}_{l_{2}} \Lambda^{J}_{t-u_{1}+u_{2},t-u_{1}+u_{3}} \cdots 
S^{k_{n}}_{l_{n}} \Lambda^{J}_{t-u_{1}+u_{n},t-u_{1}} S^{k_{n+1}}_{l_{n+1}} 
\chi^{J}_{\rm S} (t-u_{1}),  \label{NZfullcounting}
\eeqa
where we change the variables as $u_{1}=t-v, u_{2}=v_{1}-v, u_{3}=v_{2}-v,\cdots $. 

From Eq.~(\ref{NZfullcounting}), we find that for equivalent bath models such that the $n$-point bath correlation functions~(\ref{bathC}) are the same, they give the same multi-time correlation functions for the system operators~(\ref{multi-time}). 

\subsection{Poisson white noise limit}
We now substitute Eq.~(\ref{2nCPoisson}) into Eq.~(\ref{NZfullcounting}) and obtain
\beq
\partial_{t}\chi^{J}_{\rm S}(t) = \mathcal{L}^{\rm Poisson}\chi^{J}_{\rm S}(t) -i J(t)A_{\rm S}\chi^{J}_{\rm S}(t),
\label{PoissonME_J}
\eeq
where $\mathcal{L}^{\rm Poisson}$ is defined in Eq.~(\ref{PoissonME}) and we use the relation $\Lambda^{J}_{t,t}=1$. 

We conclude that, in the Poisson white noise limit, the multi-time correlation functions defined in Eq.~(\ref{multi-time}) can be calculated by solving the modified Poisson noise master equation~(\ref{PoissonME_J}).  
}

%\bibliography{ref.bib}

\begin{thebibliography}{99}

\bibitem{Weiss} U. Weiss, Quantum Dissipative Systems, (World Scientific, Singapore, 2008).

\bibitem{nonHermitian} E. J. Bergholtz, J. C. Budich, and F. K. Kunst, Exceptional topology of non-Hermitian systems, \href{https://doi.org/10.1103/RevModPhys.93.015005}{Rev. Mod. Phys. {\bf 93,} 015005 (2021).}

\bibitem{renger01} T. Renger, V. May, and O. K\"uhn, Ultrafast excitation energy transfer dynamics in photosynthetic pigment--protein complexes, \href{https://doi.org/10.1016/S0370-1573(00)00078-8}{Phys. Rep. {\bf 343,} 137 (2001).}

\bibitem{ishizaki10} A. Ishizaki, T. R. Calhoun, G. S. Schlau-Cohen, and G. R. Fleming, Quantum coherence and its interplay with protein environments in photosynthetic electronic energy transfer, \href{https://doi.org/10.1039/C003389H}{Phys. Chem. Chem. Phys. {\bf 12,} 7319 (2010).}

\bibitem{Chin12} A. W. Chin, S. F. Huelga, and M. B. Plenio, Coherence and decoherence in biological systems: principles of noise-assisted transport and the origin of long-lived coherences, \href{https://doi.org/10.1098/rsta.2011.0224}{Phil. Trans. R. Soc. A. {\bf 370,} 3638 (2012).}

\bibitem{quantum_effects_biology} M. Mohseni, Y. Omar, G. Engel, and M. Plenio (Eds.). Quantum Effects in Biology. \href{https://doi.org/10.1017/CBO9780511863189}{(Cambridge University Press, Cambridge, 2014).}

\bibitem{noiseless} P. Zanardi and M. Rasetti, Noiseless Quantum Codes, \href{https://doi.org/10.1103/PhysRevLett.79.3306}{Phys. Rev. Lett. {\bf 79,} 3306 (1997).}

\bibitem{decoherence_free1} D. A. Lidar, I. L. Chuang, and K. B. Whaley, Decoherence-Free Subspaces for Quantum Computation, \href{https://doi.org/10.1103/PhysRevLett.81.2594}{Phys. Rev. Lett. {\bf 81,} 2594 (1998).}

\bibitem{Gardiner} C. Gardiner and P. Zoller,   Quantum Noise: A Handbook of Markovian and Non-Markovian Quantum Stochastic Methods with Applications to Quantum Optics, (Springer Science \& Business Media, New York, 2004).

\bibitem{Cirac09} F. Verstraete, M. M. Wolf, and J. I. Cirac, Quantum computation and quantum-state engineering driven by dissipation, \href{https://doi.org/10.1038/nphys1342}{Nature Phys. {\bf 5,} 633 (2009).}

\bibitem{decoherence_free2} D. A. Lidar, Review of Decoherence-Free Subspaces, Noiseless Subsystems, and Dynamical Decoupling, In Quantum Information and Computation for Chemistry \href{https://doi.org/10.1002/9781118742631.ch11}{(Wiley, New York, 2014).}

\bibitem{nonMarkov} I. de Vega and D. Alonso, Dynamics of non-Markovian open quantum systems, \href{https://doi.org/10.1103/RevModPhys.89.015001}{Rev. Mod. Phys. {\bf 89,} 015001 (2017).}

\bibitem{Deffner} S. Deffner and S. Campbell, Quantum Thermodynamics: An Introduction to the Thermodynamics of Quantum Information \href{https://dx.doi.org/10.1088/2053-2571/ab21c6}{(Morgan \& Claypool, San Rafael, CA, 2019).}

\bibitem{Caldeira83} A. O. Caldeira and A. J. Leggett, Path Integral Approach to Quantum Brownian Motion, \href{https://doi.org/10.1016/0378-4371(83)90013-4}{Physica A {\bf 121,} 587 (1983).}

%Shot noise (Poisson noise)
\bibitem{Shotnoise1} Ya. M. Blanter, M. B\"{u}ttiker, Shot noise in mesoscopic conductors, \href{https://doi.org/10.1016/S0370-1573(99)00123-4}{Phys. Rep. {\bf 336,} 1, (2000).}

\bibitem{Shotnoise2} B. Huard, H. Pothier, N. O. Birge, D. Esteve, X. Waintal, and J. Ankerhold, Josephson junctions, \href{https://doi.org/10.1002/andp.200710263}{Ann. Phys. {\bf 16,} 736 (2007).}

\bibitem{Shotnoise3} T. P. Purdy, R. W. Peterson, and C. A. Regal,   Observation of Radiation Pressure Shot Noise on a Macroscopic Object, \href{https://doi.org/10.1126/science.1231282}{Science {\bf 339,} 801 (2013).}

\bibitem{1fnoise} E. Paladino, Y. M. Galperin, G. Falci, and B. L. Altshuler,   $1/f$ noise: Implications for solid-state quantum information, \href{https://doi.org/10.1103/RevModPhys.86.361}{Rev. Mod. Phys. {\bf 86,} 361 (2014)}

\revision{
\bibitem{twolevelbath1} C. M\"{u}ller, J. Lisenfeld, A. Shnirman, and S. Poletto, Interacting two-level defects as sources of fluctuating high-frequency noise in superconducting circuits, \href{https://doi.org/10.1103/PhysRevB.92.035442}{Phys. Rev. B {\bf 92,} 035442 (2015).} 

\bibitem{twolevelbath2} C. M\"{u}ller, J. H. Cole, and J. Lisenfeld, Towards understanding two-level-systems in amorphous solids: insights from quantum circuits, \href{https://doi.org/10.1088/1361-6633/ab3a7e}{Rep. Prog. Phys. {\bf 82,} 124501 (2019).}

\bibitem{twolevelbath3} I. Siddiqi, Engineering high-coherence superconducting qubits, \href{https://doi.org/10.1038/s41578-021-00370-4}{Nat. Rev. Mater. {\bf 6,} 875 (2021).}

}


%-------- tunable dissipative bath modes Ref ----------------------

\bibitem{tunable0} J. Stehlik, Y.-Y. Liu, C. Eichler, T. R. Hartke, X. Mi, M. J. Gullans, J. M. Taylor, and J. R. Petta, Double Quantum Dot Floquet Gain Medium, \href{https://doi.org/10.1103/PhysRevX.6.041027}{Phys. Rev. X {\bf 6,} 041027 (2016).}

\bibitem{tunable1} M. J. Gullans, J. M. Taylor, and J. R. Petta, Probing electron-phonon interactions in the charge-photon dynamics of cavity-coupled double quantum dots, \href{https://doi.org/10.1103/PhysRevB.97.035305}{Phys. Rev. B {\bf 97,} 035305 (2018).}

\bibitem{tunable2} T. R. Hartke, Y.-Y. Liu, M. J. Gullans, and J. R. Petta, Microwave Detection of Electron-Phonon Interactions in a Cavity-Coupled Double Quantum Dot, \href{https://doi.org/10.1103/PhysRevLett.120.097701}{Phys. Rev. Lett. {\bf 120,} 097701 (2018).}

\bibitem{tunable3} B. K. Agarwalla, M. Kulkarni, and D. Segal, Photon statistics of a double quantum dot micromaser: Quantum treatment, \href{https://doi.org/10.1103/PhysRevB.100.035412}{Phys. Rev. B {\bf 100,} 035412 (2019).}

\bibitem{tunable4} M. Raghunandan, F. Wolf, C. Ospelkaus, P. O. Schmidt, and H. Weimer, Initialization of quantum simulators by sympathetic cooling, \href{https://doi.org/10.1126/sciadv.aaw9268}{Science Advances {\bf 6,} eaaw9268 (2020).}

\bibitem{tunable5} A. Purkayastha, M. Kulkarni, and Y. N. Joglekar, Emergent $\mathcal{PT}$ symmetry in a double-quantum-dot circuit QED setup, \href{https://doi.org/10.1103/PhysRevResearch.2.043075}{Phys. Rev. Research {\bf 2,} 043075 (2020).}


\bibitem{perovskite} O. Cannelli, J. Wiktor, N. Colonna, L. Leroy, M. Puppin, C. Bacellar, I. Sadykov, F. Krieg, G. Smolentsev, M. V. Kovalenko, A. Pasquarello, M. Chergui, and G. F. Mancini, Atomic-Level Description of Thermal Fluctuations in Inorganic Lead Halide Perovskites, \href{https://doi.org/10.1021/acs.jpclett.2c00281}{The Journal of Physical Chemistry Letters, {\bf 13,} 3382 (2022).}

%----------non-Gaussian---------------

\bibitem{Hanggi} J. Shao and P. H\"{a}nggi, Decoherent Dynamics of a Two-Level System Coupled to a Sea of Spins, \href{https://doi.org/10.1103/PhysRevLett.81.5710}{Phys. Rev. Lett. {\bf 81,} 5710 (1998).}

\bibitem{spinbath} N. V. Prokof'ev and P. C. E. Stamp, Theory of the spin bath, \href{https://doi.org/10.1088/0034-4885/63/4/204}{Rep. Prog. Phys. {\bf 63,} 669 (2000).}

\bibitem{Cao1} C. Y. Hsieh and J. Cao, A unified stochastic formulation of dissipative quantum dynamics. I. Generalized hierarchical equations \href{https://doi.org/10.1063/1.5018725}{J. Chem. Phys. {\bf 148,} 014103 (2018).}

\bibitem{Cao2} C. Y. Hsieh and J. Cao, A unified stochastic formulation of dissipative quantum dynamics. II. Beyond linear response of spin baths \href{https://doi.org/10.1063/1.5018726}{J. Chem. Phys. {\bf 148,} 014104 (2018).}

\bibitem{Strunz17} V. Link and W. T. Strunz,   Stochastic Feshbach Projection for the Dynamics of Open Quantum Systems, \href{https://doi.org/10.1103/PhysRevLett.119.180401}{Phys. Rev. Lett. {\bf 119,} 180401 (2017).}

\bibitem{Strunz23} V. Link, K. Luoma, and W. T. Strunz,  Non-Markovian Quantum State Diffusion for Spin Environments, \href{https://doi.org/10.1088/1367-2630/aceff3}{New J. Phys. {\bf 25,} 093006 (2023).}

\bibitem{Pekola23} J. P. Pekola and B. Karimi,   Heat bath in a quantum circuit. \href{https://doi.org/10.48550/arXiv.2310.01246}{arXiv:2310.01246.}




%Levi-Ito
\bibitem{Gardiner09} C. Gardiner,   Stochastic Methods: A Handbook for the Natural and Social Sciences, (Springer, Berlin, 2009).

\bibitem{Applebaum} D. Applebaum,   L\'{e}vy Processes and Stochastic Calculus, 2nd ed (Cambridge University Press, Cambridge, 2011).

\bibitem{Kanazawa} K. Kanazawa,   Statistical Mechanics for Athermal Fluctuation: Non-Gaussian Noise in Physics, \href{https://doi.org/10.1007/978-981-10-6332-9}{(Springer, (2017).}


%-------- quantum-classical approach---------------

\bibitem{Kubo} R. Kubo, A Stochastic Theory of Line Shape. \href{https://doi.org/10.1002/9780470143605.ch6}{In Advances in Chemical Physics, K.E. Shuler (Ed.) (1969).}

\bibitem{Mukamel78} S. Mukamel, I. Oppenheim, and John Ross,   Statistical reduction for strongly driven simple quantum systems,  \href{https://doi.org/10.1103/PhysRevA.17.1988}{Phys. Rev. A {\bf 17,} 1988 (1978).}

\bibitem{PoissonME1} J. Luczka and M. Niemiec,   A master equation for quantum systems driven by Poisson white noise, \href{http://dx.doi.org/10.1088/0305-4470/24/17/010}{J. Phys. A {\bf 24,} L1021 (1991).}

\bibitem{Kosloff} A. Levy and R. Kosloff, Quantum Absorption Refrigerator, \href{https://doi.org/10.1103/PhysRevLett.108.070604}{Phys. Rev. Lett. {\bf 108,} 070604 (2012).}

\bibitem{PoissonME2} A. Kiely, J. G. Muga, and A. Ruschaupt,   Effect of Poisson noise on adiabatic quantum control, \href{https://doi.org/10.1103/PhysRevA.95.012115}{Phys. Rev. A {\bf 95}, 012115, (2017).}

\bibitem{PoissonME3} A. Kiely,   Exact classical noise master equations: Applications and connections, \href{http://dx.doi.org/10.1209/0295-5075/134/10001}{EPL, {\bf 134}, 10001 (2021).} 


\bibitem{classicalnoise1} A. Purkayastha and K. M\o{}lmer, Nonclassical radiation from a nonlinear oscillator driven solely by classical $1/f$ noise, \href{https://link.aps.org/doi/10.1103/PhysRevA.108.053704}{Phys. Rev. A {\bf 108,} 053704 (2023).}

\bibitem{classicalnoise2} P. Groszkowski, A. Seif, J. Koch, and A. A. Clerk, Simple master equations for describing driven systems subject to classical non-Markovian noise, \href{https://doi.org/10.22331/q-2023-04-06-972}{Quantum {\bf 7,} 972 (2023).}


\bibitem{Dicke} R. H. Dicke,   Coherence in Spontaneous Radiation Processes, \href{https://doi.org/10.1103/PhysRev.93.99}{Phys. Rev. {\bf 93,} 99 (1954).}

\bibitem{Freed75} B. Yoon, J. M. Deutch, Jack H. Freed,   A comparison of generalized cumulant and projection operator methods in spin‐relaxation theory, \href{https://doi.org/10.1063/1.430417}{J. Chem. Phys. {\bf 62,} 4687 (1975).}

\bibitem{Lindblad} G. Lindblad, On the generators of quantum dynamical semigroups, \href{https://doi.org/10.1007/BF01608499}{Commun. Math. Phys. {\bf 48,} 119 (1976).}

\bibitem{GKS} V. Gorini, A. Kossakowski, and E. C. G. Sudarshan, Completely positive dynamical semigroups of N‐level systems, \href{https://doi.org/10.1063/1.522979}{J. Math. Phys. {\bf 17,} 821 (1976).}


\bibitem{Breuer} H.-P. Breuer and F. Petruccione, The Theory of Open Quantum Systems. (Oxford University Press, 2002).





\bibitem{Nakajima} S. Nakajima,   On Quantum Theory of Transport Phenomena: Steady Diffusion, \href{https://doi.org/10.1143/PTP.20.948}{Prog. Theor. Phys. {\bf 20,} 948 (1958).}

\bibitem{Zwanzig} R. Zwanzig,   Ensemble Method in the Theory of Irreversibility, \href{https://doi.org/10.1063/1.1731409}{J. Chem. Phys. {\bf 33,} 1338 (1960).}






\bibitem{DMN} I. Bena,   Dichotomous Markov noise: exact results for out-of-equilibrium systems, \href{https://doi.org/10.1142/S0217979206034881}{Int. J. Mod. Phys. B {\bf 20,} 2825 (2006).}

\bibitem{VanDenBroeck} C. Van Den Broeck,   On the relation between white shot noise, Gaussian white noise, and the dichotomic Markov process, \href{https://doi.org/10.1007/BF01019494}{J. Stat. Phys. {\bf 31,} 467 (1983).}


\bibitem{supplement} The supplementary material includes the analytical expressions of the $2n$-point bath correlation functions, discussions about possible simulation of non-Gaussian baths by combining multiple random telegraph noise baths, and an extension of the method to calculate multi-time correlation functions of some system operators. 


\bibitem{footnote} It should be noted that the detailed balance condition does not hold for general $H_{\rm S}$ and $S^{k}$. Making the theory consistent with the detailed balance condition in a general setup is left for future work.

%pseudo modes
\bibitem{pseudo1} B. M. Garraway,   Nonperturbative decay of an atomic system in a cavity, \href{https://doi.org/10.1103/PhysRevA.55.2290}{Phys. Rev. A {\bf 55,} 2290 (1997).}

\bibitem{pseudo2} A. Imamoglu, Stochastic wave-function approach to non-Markovian systems, \href{https://doi.org/10.1103/PhysRevA.50.3650}{Phys. Rev. A {\bf 50,} 3650 (1994).}

\bibitem{pseudo3} D. Tamascelli, A. Smirne, S. F. Huelga, and M. B. Plenio, Nonperturbative Treatment of non-Markovian Dynamics of Open Quantum Systems, \href{https://doi.org/10.1103/PhysRevLett.120.030402}{Phys. Rev. Lett. {\bf 120,} 030402 (2018).}

\bibitem{pseudo4} N. Lambert, S. Ahmed, M. Cirio, and F. Nori, Modelling the ultra-strongly coupled spin-boson model with unphysical modes, \href{https://doi.org/10.1038/s41467-019-11656-1}{Nat. Commun. {\bf 10,} 3721 (2019).}


\bibitem{RC1} R. Martinazzo, B. Vacchini, K. H. Hughes, and I. Burghardt, Communication: Universal Markovian reduction of Brownian particle dynamics, \href{https://doi.org/10.1063/1.3532408}{J. Chem. Phys. {\bf 134,} 011101 (2011).}

\bibitem{RC2} A. Nazir, G. Schaller, The Reaction Coordinate Mapping in Quantum Thermodynamics, In: F. Binder, L. A. Correa, C. Gogolin, J. Anders, and G. Adesso (eds.), "Thermodynamics in the quantum regime - Recent Progress and Outlook", \href{https://doi.org/10.1007/978-3-319-99046-0_23}{(Springer International Publishing, 2018).}

\bibitem{RC3} J. Iles-Smith, N. Lambert, and A. Nazir, Environmental dynamics, correlations, and the emergence of noncanonical equilibrium states in open quantum systems, \href{https://doi.org/10.1103/PhysRevA.90.032114}{Phys. Rev. A {\bf 90,} 032114 (2014).}

\bibitem{tensor} M. Brenes, J. J. Mendoza-Arenas, A. Purkayastha, M. T. Mitchison, S. R. Clark, and J. Goold, Tensor-Network Method to Simulate Strongly Interacting Quantum Thermal Machines, \href{https://link.aps.org/doi/10.1103/PhysRevX.10.031040}{Phys. Rev. X {\bf 10,} 031040 (2020).}


%\bibitem{Neill23} N. Lambert, M. Cirio, J.-D. Lin, P. Menczel, P. Liang, F. Nori, Fixing detailed balance in ancilla-based dissipative state engineering, arXiv:2310.12539.




\bibitem{Qutip1} J. R. Johansson, P. D. Nation, and F. Nori, QuTiP: An open-source Python framework for the dynamics of open quantum systems. \href{https://doi.org/10.1016/j.cpc.2012.02.021}{Comp. Phys. Comm. {\bf 183,} 1760–1772 (2012).}

\bibitem{Qutip2} J. R. Johansson, P. D. Nation, and F. Nori, QuTiP 2: A Python framework for the dynamics of open quantum systems. \href{https://doi.org/10.1016/j.cpc.2012.11.019}{Comp. Phys. Comm. {\bf 184,} 1234 (2013).}


\end{thebibliography}
%\bibliographystyle{unsrt}

\end{document}